\documentclass[12pt]{iopart}
\usepackage{amssymb}
\usepackage{graphicx}
\usepackage{subfigure}
\usepackage{epstopdf}
\usepackage{color}
\usepackage[misc]{ifsym}
\usepackage{threeparttable}
\usepackage{xfrac}
\usepackage{algorithm}
\usepackage{algorithmicx}
\usepackage[noend]{algpseudocode}
\usepackage{dsfont}
\usepackage{subfigure}
\usepackage{booktabs}
\usepackage{multirow}

\newcommand{\ie}{\emph{i.e., }}
\newcommand{\eg}{\emph{e.g., }}

\newcommand{\mycomment}[1]{\textcolor{black}{#1}}

\begin{document}

\title[Variational quantum compiling with double $Q$-learning]{Variational quantum compiling with double $Q$-learning}

\author{Zhimin He$^{1,2}$, Lvzhou Li$^{3}$, Shenggen Zheng$^2$, Yongyao Li$^{4}$, Haozhen Situ$^{5*}$}

\address{$^1$School of Electronic and Information Engineering, Foshan University, Foshan 528000, China}
\address{$^2$Peng Cheng Laboratory, Shenzhen 518055, China}
\address{$^3$Institute of Quantum Computing and Computer Science Theory, School of Computer Science and Engineering, Sun Yat-Sen University, Guangzhou 510006, China}
\address{$^4$School of Physics and Optoelectronic Engineering, Foshan University, Foshan 528000, China}
\address{$^5$College of Mathematics and Informatics, South China Agricultural University, Guangzhou 510642, China}
\ead{situhaozhen@gmail.com}
\vspace{10pt}
\begin{indented}
\item[]
\end{indented}

\begin{abstract}
Quantum compiling aims to construct a quantum circuit $V$ by quantum gates drawn from a native gate alphabet, which is functionally equivalent to the target unitary $U$. It is a crucial stage for the running of quantum algorithms on noisy intermediate-scale quantum (NISQ) devices. However, the space for structure exploration of quantum circuit is enormous, resulting in the requirement of human expertise, hundreds of experimentations or modifications from existing quantum circuits. In this paper, we propose a variational quantum compiling (VQC) algorithm based on reinforcement learning (RL), in order to automatically design the structure of quantum circuit for VQC with no human intervention. An agent is trained to sequentially select quantum gates from the native gate alphabet and the qubits they act on by double $Q$-learning with $\epsilon$-greedy exploration strategy and experience replay. At first, the agent randomly explores a number of quantum circuits with different structures, and then iteratively discovers structures with higher performance on the learning task. Simulation results show that the proposed method can make exact compilations with less quantum gates compared to previous VQC algorithms. It can reduce the errors of quantum algorithms due to decoherence process and gate noise in NISQ devices, and enable quantum algorithms especially for complex algorithms to be executed within coherence time. 
\end{abstract}

\section{Introduction}
Current available quantum devices are not large-scale and fault-tolerant, but noisy intermediate-scale quantum\,(NISQ) devices\,\cite{PJ18}. IBM's and Rigetti's quantum devices fall into this category. \,Many quantum algorithms have been implemented by parameterized quantum circuits (PQCs) on NISQ devices.
Hybrid quantum-classical algorithms based on PQCs are successfully applied to attacking scaled-down problems in chemistry and combinatorial optimization\,\cite{OBK16,KMT17,MBB18}. Besides, they have been applied to many machine learning tasks, \eg classification\,\cite{GBC18,HCT19}, clustering\,\cite{OMA17} and generative model\,\cite{BGP19,SHW20}. These algorithms are not restricted to theory and simulation. They have been tested on real quantum devices such as IBM's and Rigetti's quantum devices.

Although NISQ device is a milestone in the field, there are several constraints, which separate them from fully utilizing their capabilities for a quantum advantage. The connectivities between qubits are limited, \ie multi-qubit gates cannot apply to arbitrary qubits. Besides, the native gate alphabet is restricted. Different quantum devices have their own topologies, \ie the arrangement of physical qubits, the interactions supported between them, and the native gate alphabet of available quantum gates. For example, IBM and Rigetti's quantum devices have different topologies.
Even different quantum devices developed by IBM (\eg 5-qubit, 16-qubit and 20-qubit devices) have different topologies.
Thus, a quantum circuit cannot directly execute on NISQ devices.

Similar as software must be complied to machine code in classical computers, quantum algorithms must be compiled before running on NISQ devices by taking into account the specific constraints.
Quantum compiling is a crucial step in the execution of quantum algorithms on NISQ devices, which receives increasing attention in industry and academia\,\cite{chong17,haner18,jurcevic2020demonstration}.
Quantum compiling aims to construct a quantum circuit by quantum gates drawn from NISQ device's native gate alphabet, which is functionally equivalent to the target unitary.
Many techniques have been proposed to successfully compile one-qubit gates\,\cite{FA11,BGS13,KMM13,YM18}. However, compiling multi-qubit unitaries is more difficult as the exponentially increasing complexity of the problem.
Hybrid quantum-classical algorithm has been applied to quantum compiling\,\cite{heya2018,jones2018,KLP19,sharma20}, which is referred to variational quantum compiling (VQC).  In VQC, a parameterized quantum circuit is trained to be functionally equivalent to the target unitary.
Heya \etal proposed a gate optimization algorithm to construct a target multi-qubit gate by parameterized single-qubit gates and multi-qubit gates\,\cite{heya2018}.
A circuit template with fixed gate types and sequence is used to compile a 7-qubit unitary with 186 gates based on energy dissipation\,\cite{jones2018}.
Circuit design is a key problem of quantum compiling, which requires human expertise and hundreds of experimentations or modifications from existing quantum circuits.
Khatri \etal proposed to perform circuit design based on simulated annealing for single-qubit and  two-qubit quantum unitaries\,\cite{KLP19}.
However, the structure of the trainable unitary is set to the same structure as the target unitary for the target unitaries with more than two qubits.
Most of current quantum compiling algorithms for large-scale gate sequences use a specific template as the searching space of the optimization over circuit structure increases exponentially with the number of quantum gates\,\cite{jones2018,KLP19,sharma20,singh19,xu19}.
However, the quantum circuits designed by this strategy might not be optimal and require more quantum gates. The performance of the quantum algorithm decreases with runtime because of the noise and limited coherence time in current NISQ devices. The increasing number of quantum gates will lead to more gate noises and increase the running time,  resulting in the performance degradation.
\mycomment{Variable structure ansatzes have been considered in \cite{cincio2018learning} to discover quantum algorithms for computing the overlap between two states, which have shorter depths than the Swap Test.}


In this paper, we proposed to automate the process of circuit design for quantum compiling based on reinforcement learning.
An agent is trained to sequentially choose quantum gates and the qubits they act on by double $Q$-learning with $\epsilon$-greedy exploration strategy\,\cite{mnih15} and experience replay\,\cite{lin93}. In $\epsilon$-greedy exploration strategy, the agent firstly performs a number of explorations by randomly generating quantum circuits,  and then slowly transforms smoothly from exploration to exploitation to generate higher performing quantum circuits according to its findings.
As the proposed VQC based on double $Q$-learning does not rely on human experience, it may have more potential to discover novel structures of circuits compared to VQC based on specific templates, which might inspire designers of quantum algorithms.
Similar work can be found in Ref.\,\cite{zhang20}, which proposed to compile single-qubit gates with elementary gates from a finite universal set based on deep reinforcement learning. However, their work is related to topological compiling and is a special case of quantum compiling by considering the action of target unitary on a fixed input state. \mycomment{In addition, reinforcement learning has also been applied for Quantum Approximate Optimization Algorithm problems \cite{khairy2019reinforcement}.}

The rest of the paper is organized as follows. Some related work is introduced in Section 2. In Section 3, we present the proposed circuit design in VQC based on double $Q$-learning. We report the performance of the proposed algorithm compared to current VQC by numerical simulation in Section 4. Finally, we summarize the results of this paper and discuss the future work in Section 5.

\section{Related work}
Variational quantum compiling (VQC) aims to compile a target unitary $U$ to a trainable quantum gate sequence $V$ which has approximately the same action as $U$ on all input states\,\cite{heya2018,KLP19,sharma20}.
Another kind of VQC only considers a specific input state, \eg $|\psi _0 \rangle=|0\rangle^{\bigotimes n}$\,\cite{jones2018}. In this paper, we focus on the former case as it has more application scenarios.

\begin{figure}
\centering
\includegraphics[width=7.9cm]{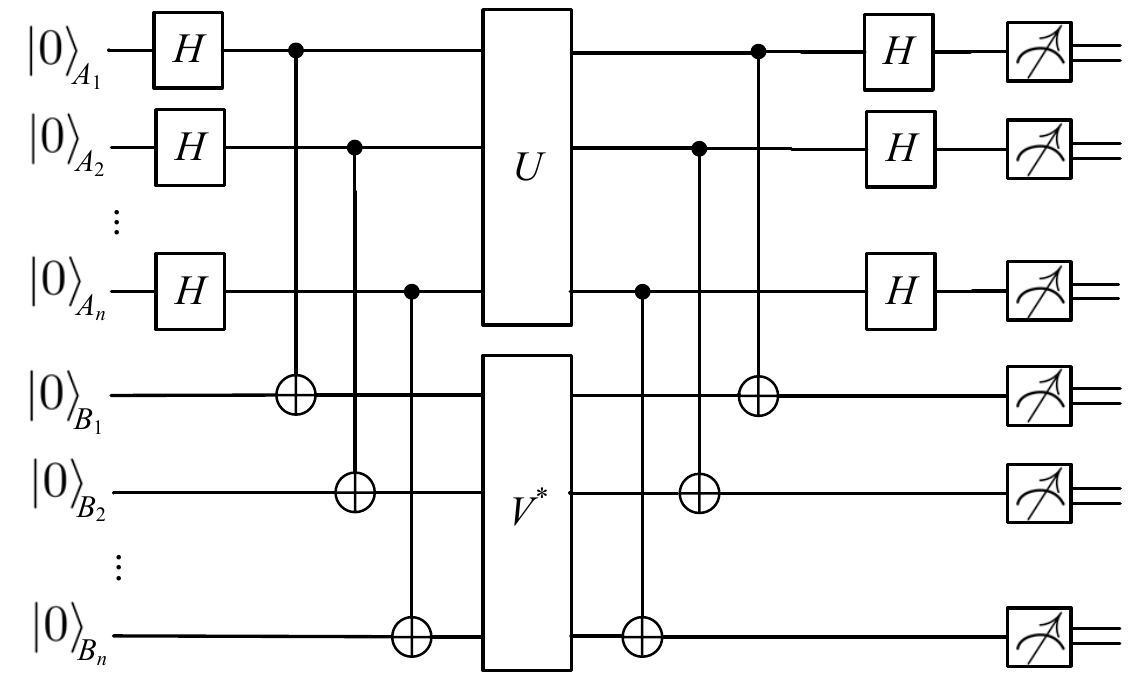}
\caption{Quantum circuit to estimate the difference between the unitaries $U$ and $V$.}
\label{fig:HSTCircuit}
\end{figure}

In VQC,  the trainable  gate sequence $V$ can be expressed in terms of the native gate alphabet of NISQ devices being used. The native gate alphabet $\mathcal{A} = \{G_k(\theta)\}_k$ consists of gates $G_k(\theta)$ that are native to the NISQ device. $k$ is a discrete parameter used to identify the type of quantum gate and the qubit it acts on. $\theta$ is the parameter of the quantum gate. If the quantum gate has no parameter, \eg Hadamard gate, then $\theta=[]$. The goal of VQC is to search a gate sequence $V_{\vec k}(\vec\theta) = G_{k_L}(\theta_L)G_{k_{L-1}}(\theta_{L-1})...G_{k_1}(\theta_1)$ with $L$ gates to minimize the difference between the target unitary $U$ and
$V_{\vec k}(\vec\theta)$, \ie

\begin{eqnarray}
(\vec{k}^{\star},\vec{\theta}^{\star})  = \arg \min_{\vec{k},\vec{\theta}} C(U,V_{\vec{k}}(\vec{\theta})),
\label{eq:obj}
\end{eqnarray}
where $\vec{k} =\{k_L,k_{L-1},...,k_1\}$ is a vector of indices describing which gates are used and which qubits they act on.  $\vec{\theta} = \{\theta_L,\theta_{L-1},...,\theta_1\}$ is a vector of continuous parameters associated with the quantum gates in $\vec{k}$.
Thus, it needs to make numerous choices to construct a quantum circuit, \ie the number of quantum gates of each type, the qubits they act on, the ordering of quantum gates, and the parameters of quantum gates, which makes the space of quantum circuit structures extremely large and infeasible for  an exhaustive manual search.
$C(U,V_{\vec{k}}(\vec{\theta})$ quantifies how close the trained unitary $V_{\vec{k}}(\vec{\theta})$ is to the target unitary $U$. It can be calculated by the Hilbert-Schmidt inner product between unitaries $U$ and $V$\,\cite{KLP19}, \ie
\begin{eqnarray}
C(U,V) = 1-\frac{|\langle V, U \rangle|^2}{d^2}=1-\frac{|\text{Tr}( V^\dagger U)|^2}{d^2},
\end{eqnarray}
where $d=2^n$ is the Hilbert-space dimension where $n$ is the number of qubits in $U$. $C(U,V) = 0$ if and only if unitary $U$ and unitary $V$ differ by a global phase factor. $C(U,V)$ can be calculated by the quantum circuit in Fig.\,\ref{fig:HSTCircuit}\,\cite{KLP19}.
At the beginning of the circuit, a maximally-entangled state between $n$-qubit systems $A$ and $B$ is prepared with $n$ Hadamard gates and $n$ CNOT gates. Then the unitaries $U$ and $V^*$ act on system A and B, respectively, where $V^*$ is the complex conjugate of
unitary $V$.  The undoing unitary is implemented with $n$ Hadamard gates and $n$ CNOT gates. At last, $C(U,V)$ is calculated by the probability of the all-zeros outcome for the Bell-basis measurement.
\mycomment{For large problem sizes, the global Hilbert-Schimdt Test exhibits Barren Plateau phenomena\,\cite{mcclean2018barren,cerezo2020cost}. In this paper, local Hilbert-Schmidt Test\,\cite{KLP19} is used for the large problem sizes to address this problem:
\begin{eqnarray}
C_{local}(U,V) = \frac{1}{n}\sum_{i=1}^{n}C_{local}^{(i)}(U,V),
\end{eqnarray}
where $C_{local}^{(i)}(U,V)$ is the same as the global Hilbert-Schmidt Test $C(U,V)$ except that only qubits $i$ and $i+n$ are measured, \ie the probability of the all-zeros outcome on qubits $i$ and $i+n$  for the Bell-basis measurement. $C_{local}(U,V)$ is efficient to compute on a quantum computer and scales well with the size of the problem \cite{KLP19}.}

The optimization over the continuous parameters $\vec{\theta}$ can be performed by a gradient descent algorithm\,\cite{KLP19}.
The gradient with respect to  $\theta_l$ of $C(U,V_{\vec{k}}(\vec{\theta}))$ is
\begin{eqnarray}
\frac{\partial C(U,V_{\vec{k}}(\vec{\theta})}{\partial \theta_l} = \frac{1}{2}(C(U,V_{\vec{k}}(\vec{\theta}^+_l)-C(U,V_{\vec{k}}(\vec{\theta}^-_l),
\label{eq:par}
\end{eqnarray}
where $\vec{\theta}^\pm_l=\vec{\theta}\pm \frac{\pi}{2}\vec{e}^{\,l}$, and $\vec{e}^{\,l}$ is the $l$th unit vector in the  parameter space corresponding to
$\theta_l$, \ie $\theta_l \leftarrow \theta_l \pm \frac{\pi}{2}$, with other angles $\vec{\theta} - \{\theta_l\}$ unchanged.
\mycomment{The derivation of Eq. (\ref{eq:par}) adopts the parameter shift rule\,\cite{schuld2019evaluating}, which does not need any additional quantum gates and has been widely used in many hybrid quantum-classical algorithms\,\cite{SHW20,KLP19,sharma20}. Note that for the parameter shift rule to be valid, the generator of the parameterized gate needs to have two unique eigenvalues.}

The calculation of  $\vec k $ in Eq.\,(\ref{eq:obj}) is an optimization over gate structures. Many current quantum compiling algorithms use a specific template\,\cite{jones2018,KLP19,sharma20,singh19}.
Simulated annealing is also used for the optimization over gate structure\,\cite{sharma20,cincio20}. However, the searching space of gate structures increases exponentially with the number of gates, \mycomment{\ie the total number of quantum gates in the circuit}, where simulated annealing is less efficient. In this paper, we focus on the optimization over gate structure and propose to find an optimal gate structure by reinforcement learning.

\section{Design of quantum circuits with double $Q$-learning}
Reinforcement learning aims to make optimal decision using experiences. $Q$-learning is a model-free reinforcement learning algorithm incorporating $\epsilon$-greedy strategy and experience replay, which has been proved powerful in large state space\,\cite{AB11,VM05,Baker17,zhong20}.
The first AI program to beat a professional human Go player, \ie AlphaGO, is fundamentally $Q$-learning. In this paper, we adopt $Q$-learning to automatically design a quantum circuit on the large space of quantum circuit structures.
$Q$-learning maps \{state, action\} pair to $Q$-value, which is an estimation of how good it takes the action at the state.
The fundamental thing of $Q$-learning is to iteratively update the $Q$-values of \{state, action\} pairs based on the immediate reward and the future reward after taking the action.

\begin{figure*}
\centering
\subfigure[]{
\includegraphics[width=0.99\textwidth]{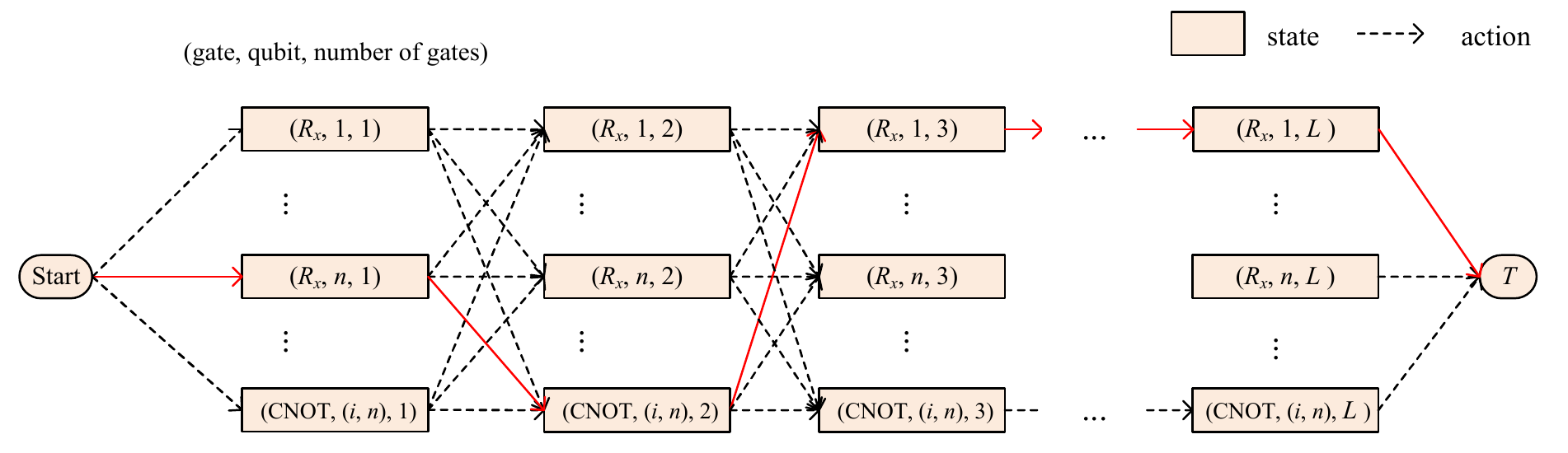}}
\subfigure[]{
\includegraphics[width=0.45\textwidth]{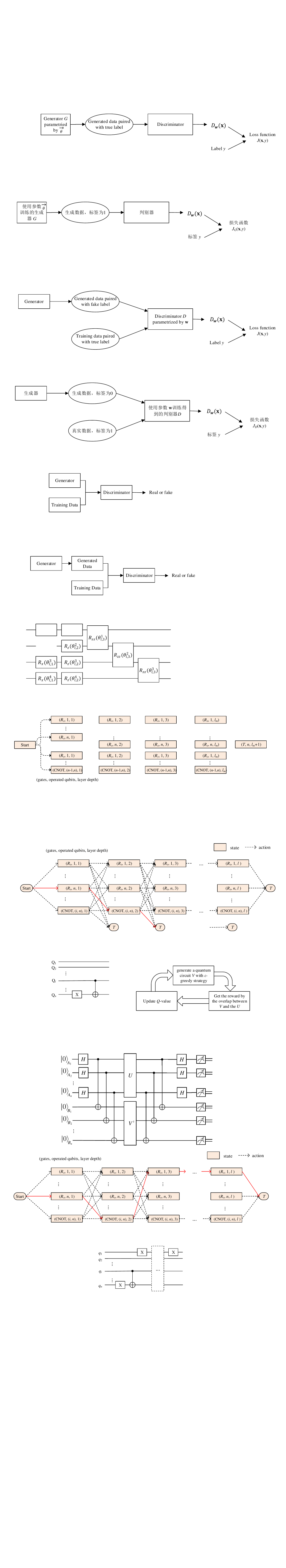}}
\subfigure[]{
\includegraphics[width=0.5\textwidth]{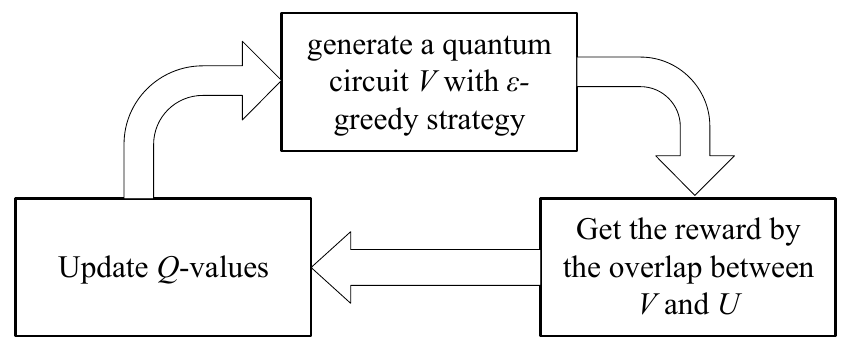}}
\caption{Illustration of quantum circuit construction with $Q$-learning. (a) shows the feasible states, action space of a $n$-qubit quantum circuit with $L$ quantum gates and the state transitions by different actions.
The tuple of each state is (gate, qubit, number of gates). For example, ($R_x$,$n$,3) denotes that the quantum circuit consists of  three gates and the last gate of the quantum circuit is $R_x$ acting on the $n$th qubit. $T$ state denotes a termination state. The red solid line shows a path the agent may choose, resulting on the quantum circuit in (b). The flowchart of $Q$-learning is shown in (c).}
\vspace{-10pt}
\label{Fig:transitionFlow}
\end{figure*}

In this paper, we train a learning agent to sequentially choose quantum gates from the native gate alphabet until a termination state as described in Fig.~\ref{Fig:transitionFlow}(a).
Each state is defined as a tuple of all relevant parameters, \ie (gate, qubit, number of gates), \mycomment{which denote the type of the selected quantum gate, the qubit the gate acts on and the number of gates in current gate sequence}.
In order to make sure the state-action graph is directed and acyclic, the parameter \emph{number of gates} is added to the state space. Another advantage of equipping the state with \emph{number of gates} is that it can limit the number of quantum gates in the compiled quantum circuit. Any state with the maximum number of gates can only transition to a termination state.
The selection of a quantum gate and the qubit it acts on can be regarded as an action in $Q$-learning.
The number of actions in the current state depends on the native gate alphabet and the number of qubits. For example, without considering the connectivity constraint, the number of actions is $np+qn(n-1)$ for compiling an $n$-qubit unitary with the native gate alphabet including $p$ one-qubit gates and $q$ two-qubit gates.
Some restrictions should be considered to make the learning processing tractable. At first, the state with $j$ gates is only permitted to transition to a state with $j+1$ gates, which ensures that the state-action graph is directed and acyclic.
Fig.\,\ref{Fig:transitionFlow}(a) shows feasible states, action space and the state transition from $s_k$ to $s_{k+1}$ by taking action $a_k$, where $k$ is the number of gates in current state.

The quantum circuit in Fig.\,\ref{Fig:transitionFlow}(b) is generated by the red solid line in (a).
The gate selection process is a Markov decision process. The design of a quantum circuit can be regarded as an action selection trajectory $\tau_{i}$.
The reward $r$ of the transition from state $s$ to $s'$ by action $a$ may be stochastic according to a distribution $p_{r|s,a,s'}$.
The agent aims to maximize the expected reward over all possible trajectories $\tau$, \ie $\max_{\tau_{i}\in \tau } R_{\tau_{i}}$. The total expected reward for a trajectory $\tau_{i}$ is
\begin{eqnarray}
R_{\tau_{i}} = \sum _{(s,a,s')\in \tau_{i}}\mathbb{E}_{r|s,a,s'}[r|s,a,s'].
\label{eq:re}
\end{eqnarray}
Given a state $s_t\in S$ and subsequent action $a\in A(s_t)$ where $S$ is all possible states and $A(s_t)$ is all possible actions in state $s_t$, the maximum total expected reward is denoted by $Q^{\star}(s_t,a)$, i.e., $Q$-value of the state-action pair $(s_t, a)$. Based on Bellman's equation, we can get
\begin{eqnarray}
Q^{\star}(s_t,a) = \mathbb{E}_{r|s_t,a,s_{t+1}}[r|s_t,a,s_{t+1}]+\gamma \max_{a' \in A(s_{t+1})}Q^{\star}(s_{t+1},a')
\label{eq:bellEq}
\end{eqnarray}
It can be formulated as an iterative update
\begin{eqnarray}
Q(s_t,a) = (1-\alpha)Q(s_t,a) + \alpha( r_{t}+\gamma Q(s_{t+1},a^{\star})),
\label{eq:QIter}
\end{eqnarray}
where $a^{\star} = \arg \max_{a' \in A(s_{t+1})}Q(s_{t+1},a')$. $\alpha \in [0,1]$ is the learning rate, which determines the importance of the newly acquired information compared to old information.
$\alpha=0$ means that the agent learns nothing from the newly acquired information, while  $\alpha=1$ indicates the agent only considers the most recent information ignoring prior knowledge.
$\gamma$ is the discount factor, which makes a balance between the immediate reward and the future reward. $\gamma=0$ means that the agent only considers immediate reward. The future reward is more important when $\gamma$ is larger.
$r_{t}$ is the intermediate reward when moving from state $s_t$ to $s_{t+1}$.
When a state goes from $s_t$ to a termination state $s_T$ by an action $a_T$, Eq.\,(\ref{eq:QIter}) can be simplified as
\begin{eqnarray}
Q(s_t,a_T) = (1-\alpha)Q(s_t,a_T) + \alpha r_{T},
\label{eq:QIterT}
\end{eqnarray}
where $r_T$ is the termination reward. It can be calculated by
\begin{eqnarray}
r_T = 1 - C(U,V ),
\label{eq:r}
\end{eqnarray}
where $C(U,V )$ is measured by Hilbert-Schmidt Test of current quantum circuit and the target unitary $U$ using the circuit in Fig.\,\ref{fig:HSTCircuit}.
The intermediate reward $r_t$ can be estimated by reward shaping\,\cite{NHR99}, which can speed up the training process, \ie $r_t = r_T/L$, where $L$ is the number of gates in the quantum circuit.

The detail process of double $Q$-learning for circuit design in VQC is shown in Algorithm\,\ref{alg:qlearning}.
At first, $Q$-values are initialized by the average of the termination reward $r_T$ over a number of randomly generated quantum circuits, \ie the quantum circuits generated when $\epsilon = 1$.
For each $\epsilon_j \in \vec \epsilon$, $E_j$ circuits are generated based on $\epsilon$-greedy strategy.
The parameters $\vec \theta$ of these quantum circuits are optimized based on the gradient descent approach in Algorithm 4 of Ref.~\cite{KLP19}.
Then the reward can be calculated by Hilbert-Schmidt Test with the quantum circuit in Fig.\,\ref{fig:HSTCircuit}, and is stored in a replay memory as well as its corresponding circuit structure.
The $Q$-values are periodically updated by a batch samples randomly drawn from the replay memory.
The flowchart of $Q$-learning is shown in Fig.\,\ref{Fig:transitionFlow}(c).
The performance of $Q$-learning is not satisfying in stochastic Markov decision processes due to the overestimation of the action values resulted from using the maximum action value as an approximation for the maximum expected action value.
In this paper, we use double $Q$-learning\,\cite{van16}, which uses two $Q$-value functions $Q_1$ and  $Q_2$ to determine the value of the next state.
In $\epsilon$-greedy strategy, a quantum gate is selected randomly with probability $\epsilon$ and based on $Q$-value with probability $1-\epsilon$. At first, with a high $\epsilon$, the agent can explore different structures of quantum circuits by randomly choosing actions. As the agent explores the environment, $\epsilon$ decreases and the agent starts to exploit the environment.

\begin{algorithm}[t]
\renewcommand{\algorithmicrequire}{\textbf{Input:}}
\renewcommand\algorithmicensure {\textbf{Output:} }
\caption{Double $Q$-learning for circuit design in VQC}
\label{alg:qlearning}
\begin{algorithmic}[1]
\Require
$\vec \epsilon=\{\epsilon_1,\epsilon_2,...,\epsilon_m\}$: parameters of $\epsilon$-greedy strategy;
$E_j$: the number of generated quantum circuits for $\epsilon_j$;
$K$: the size of minibatch for $Q$-value update;
$L$: the maximum gate number of the quantum circuit;
$\gamma$: the discount factor;
$\alpha$: the learning rate;
$\mathcal{A}$: the native gate alphabet.

\Ensure $V^{\star}$: the compiled quantum circuit.
\State set the replay memory $M=[]$ and set $r^{\star} = 0$
\State set $Q_1(s,a)=q_0, Q_2(s,a)=q_0, \forall s\in S, \forall a\in A(s)$, where $q_0$ is the average of the termination reward $r_T$ over a number of randomly generated quantum circuits and the action space depends on the native gate alphabet $\mathcal{A}$
\For{each $\epsilon_j$ in $\vec \epsilon$}
    \For{\texttt{$i$ = 1 to $E_j$}}
            \State remark: generate the $i$th quantum circuit with $\epsilon$-greedy strategy
            \State set gate sequence $V_i=[]$ and current state $s$ = `Start'
            \For{\texttt{$l$ = 1 to $L$}}
                \State with probability $\epsilon_j$ select a random action $a'$ from $A(s)$
                \State otherwise select action $a' = \arg\max_{a\in A(s)}(Q_1(s,a)+Q_2(s,a))$
                \State perform a transition($s = s'$) from current state to $s'$ after taking action $a'$
                \State append $(s,a')$ to ${V}_i$
            \EndFor
            \State generate a quantum circuit based on $V_i$
           \State optimize the gate parameters $\vec \theta$ based on the gradient descent approach in\,\cite{KLP19}
            \State calculate the reward $r_{T}^{i}$ according to Eq.\,(\ref{eq:r})
            \State put $\{V_i,r_{T}^i  \}$ into the replay memory $M$
            \If {$r^{\star} < r_{T}^{i}$}
                \State $r^{\star} = r_{T}^{i}$ and $V^{\star} = V_i$
            \EndIf
            \State remark: update $Q$-values by a batch samples drawn from the replay memory
            \For {\texttt{$k=1$ to $K$}}
                \State randomly select $\{V_k,r_{T}^k\}$ from the replay memory
                \State generate a random number $y$ between 0 and 1
                \For{\texttt{each $(s_t,a_t)$  in $V_k$}}
                    \If {$y<0.5$}
                        \State$Q_1(s_t,a_t) = (1-\alpha)Q_1(s_t,a_t) + \alpha( r_{t}+\gamma Q_2(s_{t+1},a^{\star})),$
                        \State where $a^{\star}= \arg \max_{a \in A(s_{t+1})}Q_1(s_{t+1},a)$
                    \Else
                        \State$Q_2(s_t,a_t) = (1-\alpha)Q_2(s_t,a_t) + \alpha( r_{t}+\gamma Q_1(s_{t+1},a^{\star})),$
                        \State where $a^{\star}= \arg \max_{a \in A(s_{t+1})}Q_2(s_{t+1},a)$
                    \EndIf
                \EndFor
            \EndFor
    \EndFor
\EndFor
\end{algorithmic}
\end{algorithm}

\mycomment{As two-qubit gates have much lower fidelities than single-qubit gates in current NISQ devices, it is of great significance to minimize the number of two-qubit gates in the compiled circuit \cite{BKM14,gaebler16}. It can be done by modifying the reward function in Eq.\,(\ref{eq:r}) as
\begin{eqnarray}
r_T = 1 - C(U,V)-\lambda*C_p,
\label{eq:rCNOT}
\end{eqnarray}
where $C_p = n_{\text{CNOT}}/L$ is a penalty term which motivates the agent to use fewer CNOT gates in the designed circuit. $\lambda$ is a weighted parameter which balances $C(U,V)$ and the number of CNOT gates in the circuit. $n_{\text{CNOT}}$ and $L$ are the number of CNOT gates and quantum gates in the circuit.}

\mycomment{The searching space of the optimization over circuit structure increases exponentially with the number of quantum gates  $L$. In many applications, $L$ grows polynomially with the number of qubits. This issue is crucial for large problem sizes. To address this problem, we can add some common blocks to the candidate gate alphabet, which largely decreases the computation complexity of the proposed algorithm for large problem sizes. This strategy can be regarded as a crossover of variable structure ansatze and fixed-template compiling, which not only makes the algorithm scale well with the size of the problem, but also decreases the number of quantum gates in the compiled circuit.}

\section{Numerical simulation}
In this section, we verify the performance of the proposed variational quantum compiling (VQC) based on double $Q$-learning on 2 to 8-qubit unitaries.
\subsection{VQC on two-qubit unitaries}
In this section, we compile five two-qubit unitaries including controlled-phase\,(CS), controlled-Hadamard\,(CH), controlled-Z\,(CZ), M{\o}lmer-S{\o}rensen $XX$ gate with $\theta=3\pi/2$\,($XX(3\pi/2)$) and two-qubit quantum Fourier transform\,(QFT2) in a simulator. Similar to the previous work\,\cite{KLP19}, we consider the native gate alphabet
\begin{eqnarray}
\mathcal{A} = \{R_x(\pi/2), R_z(\theta), \text{CNOT}\},
\end{eqnarray}
where $R_x(\pi/2)$ and $R_z(\theta)$ are one-qubit gates while CNOT is a two-qubit gate. $\theta$ is a continuous parameter of $R_z$ gate which denotes the angle to rotate around the z-axis on the Bloch sphere. In this simulation, we do not consider the connectivity constraint of NISQ devices. Thus, the number  of actions ($|A(s)|$) for each state $s$ is $6$.
We can simply implement quantum compiling for NISQ devices with connectivity constraint by modifying the action space $A(s)$, \ie removing the actions which are not permitted due to the connectivity constraint.

\mycomment{A replay memory is used to store the circuit structures and their corresponding costs $C(U,V)$ after each iteration in the experience replay. The parameter $K$ indicates the agent samples $K$ quantum circuits with their corresponding costs from the replay memory to update the $Q$-values. The learning rate $\alpha$, the discount factor $\gamma$ and the batch size $K$ are empirical parameters. According to previous works\,\cite{zhong20,sutton2018reinforcement} and the results of serval values, we set $\alpha$, $\gamma$  and $K$ to be 0.02, 0.9 and 128.}

In $\epsilon$-greedy strategy, the agent either selects a quantum gate based on $Q$-values with probability $1-\epsilon$ or randomly selects a quantum gate  with probability $\epsilon$. In our experiment, $\epsilon$ decreases from 1.0 to 0.1, which makes the agent transform smoothly from exploration to exploitation. The number of quantum circuits generated for each $\epsilon$ is shown in Table \ref{Tab:numQC2Qb}. A large number of randomly generated quantum circuits are trained at $\epsilon=1.0$ in order to let the agent see more quantum circuit structures.
\begin{table}
	\centering
	\caption{Number of quantum circuits generated at each $\epsilon$ for two-qubit unitaries.}
	\begin{tabular}{ccccccccccc}
		\toprule
		$\epsilon$&1.0&0.9&0.8&0.7&0.6&0.5&0.4&0.3&0.2&0.1\\
		\midrule
		number of circuits&1500&100&100&100&150&150&150&150&150&150\\
		\bottomrule
	\end{tabular}
\label{Tab:numQC2Qb}
\end{table}

\begin{figure}
\centering
\includegraphics[width=9cm]{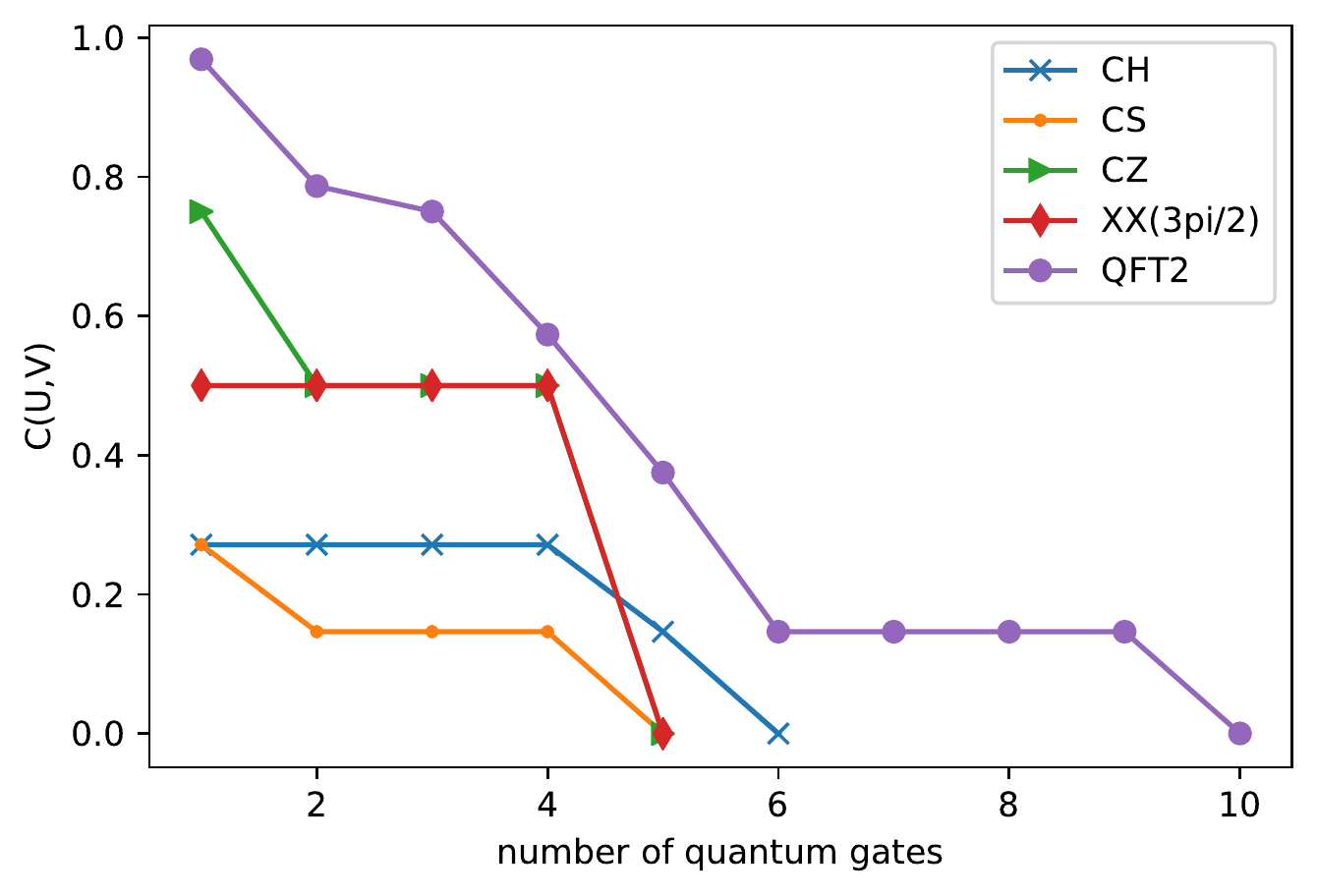}
\caption{The minimum cost achieved by VQC based on double $Q$-learning for the two-qubit unitares, \ie controlled-phase (CS), controlled-Hadamard (CH), controlled-Z (CZ), M{\o}lmer-S{\o}rensen $XX$ gate with $3\pi/2$ ($XX(3\pi/2)$) and two-qubit quantum Fourier transform (QFT2).}
\label{fig:twoQubit}
\end{figure}

Fig.~\ref{fig:twoQubit} shows the minimum cost achieved by VQC based on double $Q$-learning as the increasing of the number of gates in the quantum circuit. The cost $C(U,V)$ decreases with the increasing number of quantum gates in $V$ and finally converges to 0, \ie $U=V$.
For CS, CZ and $XX(3\pi/2)$ gates, a quantum circuit with 5 quantum gates from the native gate alphabet can achieve zero cost.
To compile CH gate, it needs at least 6 quantum gates. QFT2 is the most difficult unitary to compile which needs 10 quantum gates. \mycomment{The numbers of CNOT gates in the compiled circuits for CH, CS, CZ, $XX(3\pi/2)$ and QFT2 are 1, 2, 2, 2 and 3, respectively. We remark that the theoretical lower bound of CNOT gates required for the implementation of an arbitrary two-qubit unitary is 3 \cite{shende2004,shende2006}.}

\begin{figure}
\centering
\subfigure[CS]{\includegraphics[width=4.1cm]{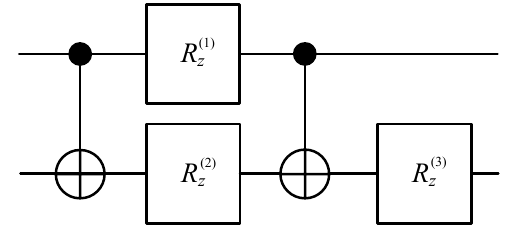}}
\subfigure[CH]{\includegraphics[width=6.1cm]{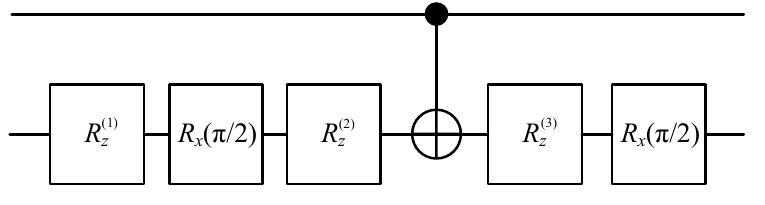}}
\subfigure[CZ]{\includegraphics[width=3.9cm]{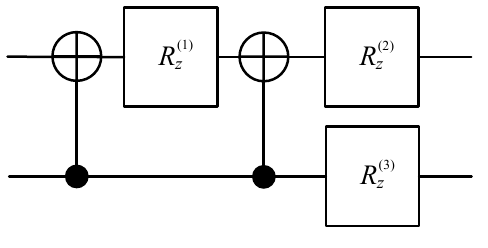}}
\subfigure[$XX(3\pi/2)$]{\includegraphics[width=3.95cm]{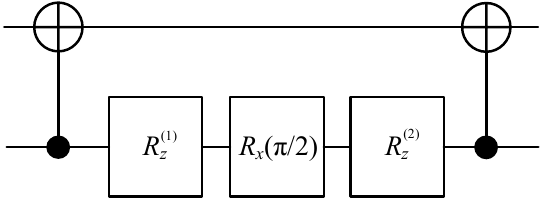}}
\subfigure[QFT2]{\includegraphics[width=6.9cm]{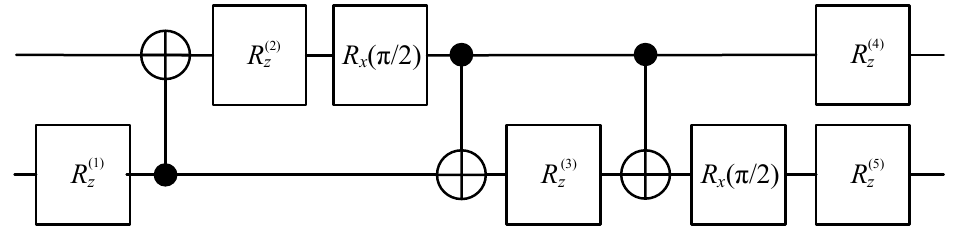}}
\caption{The compiled circuits by VQC based on double $Q$-learning for the two-qubit unitares.}
\label{fig:twoQubitCircuit}
\end{figure}

We also compare the number of quantum gates in the VQC based on double $Q$-learning to the one based on simulated annealing~\cite{KLP19} in Table~\ref{Tab:numQG2Qb} and find that we achieve an exact compilation with the same number of quantum gates. Fig.\,\ref{fig:twoQubitCircuit} shows the structures of the compiled circuits by VQC based on double $Q$-learning for the two-qubit unitares. The rotation angles of the quantum gates in these circuits are illustrated in Table \ref{Tab:par2qubit} in \ref{sec:appendix}.
\begin{table}
	\centering
	\caption{The number of quantum gates in the quantum circuits compiled by the VQC in Ref.\,\cite{KLP19} (QAQC) and the proposed VQC for the two-qubit target unitaries.}
    \begin{threeparttable}	
    \begin{tabular}{cccccc}
		\toprule
		&CS&CH&CZ&$XX(3\pi/2)$&QFT2\\
		\midrule
		QAQC&/&6&5&/&10\\
        proposed VQC &5&6&5&5&10\\
		\bottomrule
	\end{tabular}
    \begin{tablenotes}
        \footnotesize
        \item[] ``/'' means that Ref.\,\cite{KLP19} did not consider this quantum gate.
      \end{tablenotes}
    \end{threeparttable}

\label{Tab:numQG2Qb}
\end{table}

\subsection{VQC on three-qubit unitaries}
\mycomment{We firstly perform a simulation of the proposed method on fully-connected qubits in Section \ref{sec:421}. Then the limited connectivity of qubits on IBM's 5-qubit quantum device (`Ibmq$\_$ourense') is considered in Section \ref{sec:422}. In Section \ref{sec:423}, a penalty term is considered in the reward function to motivate the agent to generate quantum circuits with fewer CNOT gates.}

\subsubsection{Simulation on fully-connected qubits}
\label{sec:421}

Similar to the previous work\,\cite{sharma20}, we consider the following three-qubit unitaries: three-qubit W-state preparation (WSP3), Toffoli gate (CCNOT) and three-qubit quantum Fourier transform (QFT3). We use the same native gate alphabet as the previous work\,\cite{sharma20}, \ie
\begin{eqnarray}
\mathcal{A} = \{R_z(\theta), R_y(\theta), \text{CNOT}\},
\label{eq:gateSet}
\end{eqnarray}
where $R_x(\theta)$ and $R_y(\theta)$ are one-qubit rotation gates around x-axis and y-axis.

\begin{figure}
\centering
\includegraphics[width=9cm]{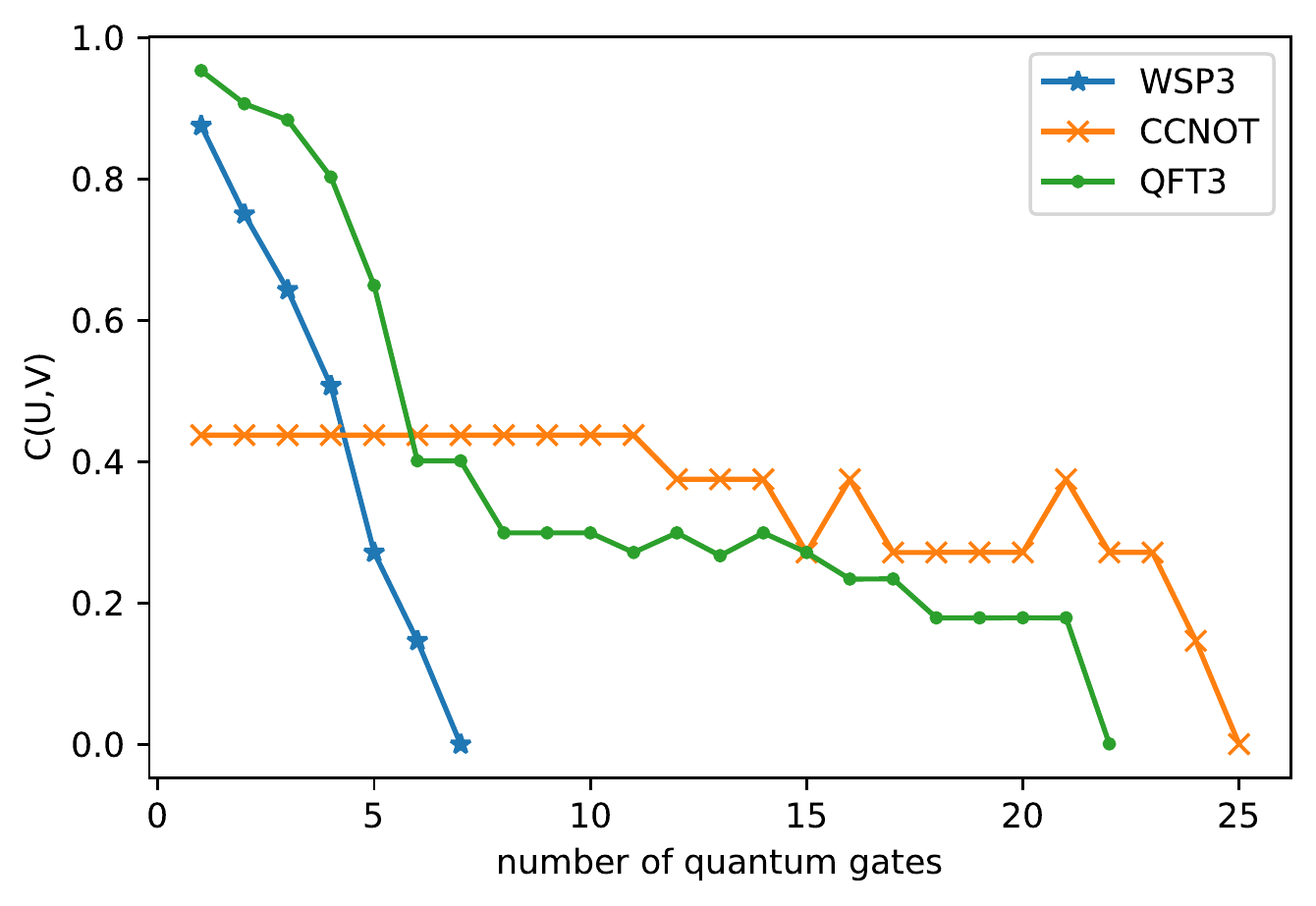}
\caption{The minimum cost achieved by VQC based on double $Q$-learning for the three-qubit unitaries, \ie three-qubit W-state preparation (WSP3), Toffoli gate (CCNOT), and three-qubit quantum Fourier transform (QFT3).}
\label{fig:threeQubit}
\end{figure}

Similar to the previous section, the learning rate $\alpha$, the discount factor $\gamma$ and the batch size $K$ are set to 0.02, 0.9 and 128, respectively.
Compared to two-qubit unitaries, it is much more difficult to compile three-qubit unitaries as it needs more quantum gates and the searching space of the optimization over gate structures increases exponentially with the number of gates. Thus, we enlarge the number of quantum circuits in each $\epsilon$ as shown in Table \ref{Tab:numQC3Qb}.
Fig.\,\ref{fig:threeQubit} shows the result of proposed VQC with different numbers of quantum gates in $V$ for different target unitaries, \ie WSP3, CCNOT and QFT3.
The cost $C(U,V)$ decreases with the increasing number of quantum gates in the circuit $V$ and converges to 0 for all the target unitaries.
WSP3 is the easiest to compile among the three target unitaries, which needs 7 quantum gates to make an exact compilation.
QFT3 and CCNOT are more difficult to compile which need 22 and 25 quantum gates, respectively. \mycomment{The numbers of CNOT gates in the compiled circuits for WSP3, CCNOT and QFT3 are 2, 10 and 7, respectively. We remark that the theoretical lower bound of CNOT gates required for the implementation of an arbitrary three-qubit unitary is 14 \cite{shende2004,shende2006}.}

\begin{table}
	\centering
	\caption{The number of quantum circuits generated at each $\epsilon$ for three-qubit unitaries.}
	\begin{tabular}{ccccccccccc}
		\toprule
		$\epsilon$&1.0&0.9&0.8&0.7&0.6&0.5&0.4&0.3&0.2&0.1\\
		\midrule
		number of circuits&3000&200&200&200&300&300&300&300&300&300\\
		\bottomrule
	\end{tabular}
\label{Tab:numQC3Qb}
\end{table}

\begin{figure*}
\centering
\subfigure[]{
\includegraphics[width=0.46\textwidth]{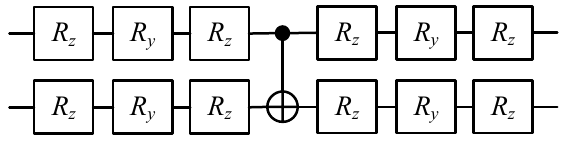}}\\
\subfigure[]{
\includegraphics[width=0.35\textwidth]{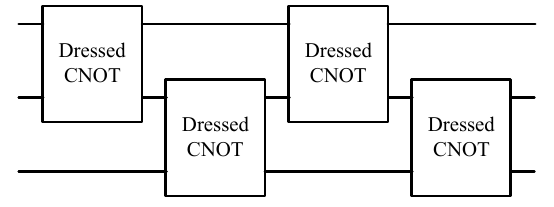}}
\subfigure[]{
\includegraphics[width=0.36\textwidth]{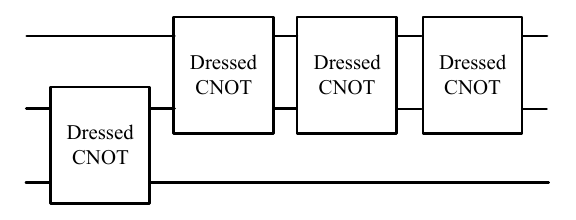}}
\caption{The structures of quantum circuits in alternating-pair ansatz with two layers and the target-inspired ansatz. (a) dressed CNOT; (b) alternating-pair ansatz with two layers; (c) target-inspired ansatz for three-qubit W-state preparation.}
\vspace{-10pt}
\label{Fig:target3Qubit}
\end{figure*}

We also compare the performance of the circuit structure designed by the proposed method and the found structures in Ref.\,\cite{sharma20}, \ie alternating-pair ansatz with two layers and the target-inspired ansatz.
In each layer of alternating-pair ansatz, dressed CNOT gates (see Fig.~\ref{Fig:target3Qubit}(a)) act on alternating pairs of neighboring qubits as described in Fig.\,\ref{Fig:target3Qubit}(b).
Target-inspired ansatz is constructed by removing all one-qubit gates and replacing each remaining gate with a dressed CNOT in the gate sequence transformed by IBM's simulator. Due to the limitation of space, we just illustrate an example of the target-inspired ansatz for WSP3 in Fig.\,\ref{Fig:target3Qubit}(c). The structures of target-inspired ansatz for CCNOT and QFT3 can be generated in the same manage.

Table \ref{Tab:comparedMethod} shows the cost achieved by alternating-pair ansatz with two layers and the target-inspired ansatz.
For WSP3, both alternating-pair ansatz and target-inspired  ansatzs can achieve a cost close to 0. However, they use 52 quantum gates while the proposed VQC only needs 7 quantum gates.
25 quantum gates are used to make an exact compilation for CCNOT in the proposed method.
Although alternating-pair ansatz uses twice as many quantum gates, \ie 52 gates, it cannot make an exact compilation.
Target-inspired ansatz can achieve a low cost close to zero. However, it uses three times as many gates as the proposed method. Besides, it needs to have knowledge of the gate sequences of the target unitaries transformed by IBM's simulator.
For QFT3,  22 quantum gates are used by our method to make an exact compilation, which is much fewer than the other two methods.
Alternating-pair ansatz cannot make an exact compilation although it uses 52 quantum gates.
The cost achieved by target-inspired ansatz is higher than the proposed method although it uses more than three times as many quantum gates as the proposed method.
The simulation results show the importance of the optimization over gate sequence in quantum compiling. The proposed VQC based on reinforcement learning can make an exact compilation with fewer quantum gates, which reduces the errors of quantum algorithm due to decoherence process and gate noise in NISQ devices, and enables quantum algorithms especially for complex algorithms to be executed within coherence time.
\begin{table}
	\centering
	\caption{The cost $C(U,V)$ achieved by the proposed VQC and the structures in \cite{sharma20}, \ie alternating-pair ansatz with two layers and the target-inspired ansatz.}
    \begin{tabular}{cccc}
    \hline
   &structure&number of gates&$C(U,V)$ \\
    \hline \multirow{3}{*}{WSP3} &  alternating-pair ansatz& 52& 1.86E-03\\
    &target-inspired ansatz&52&8.96E-04\\
    &proposed VQC&7&0\\
\hline
    \multirow{3}{*}{CCNOT} &  alternating-pair ansatz& 52& 1.47E-01\\
    &target-inspired ansatz&78&2.59E-03\\
    &proposed VQC&25&2.98E-04\\
\hline
    \multirow{3}{*}{QFT3} &  alternating-pair ansatz& 52& 3.88E-02\\
    &target-inspired ansatz&78&3.03E-03\\
    &proposed VQC&22&6.30E-04\\
    \hline
    \end{tabular}
\label{Tab:comparedMethod}
\end{table}

\begin{figure}
\centering
\subfigure[WSP3]{\includegraphics[width=3.3cm]{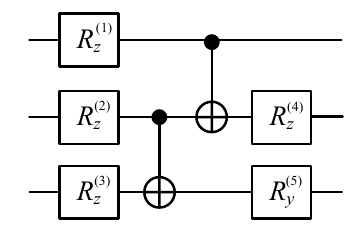}}
\subfigure[CCNOT]{\includegraphics[width=11cm]{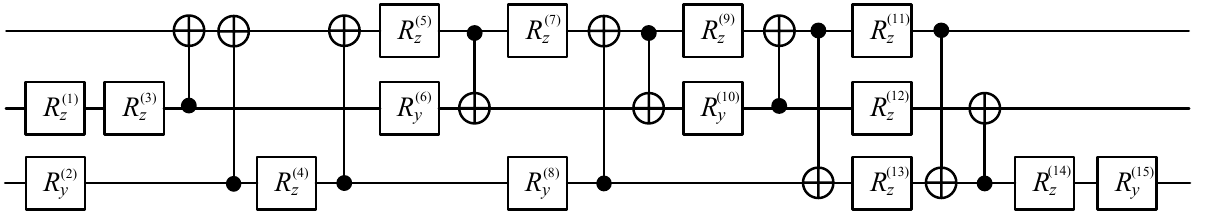}}
\subfigure[QFT3]{\includegraphics[width=12cm]{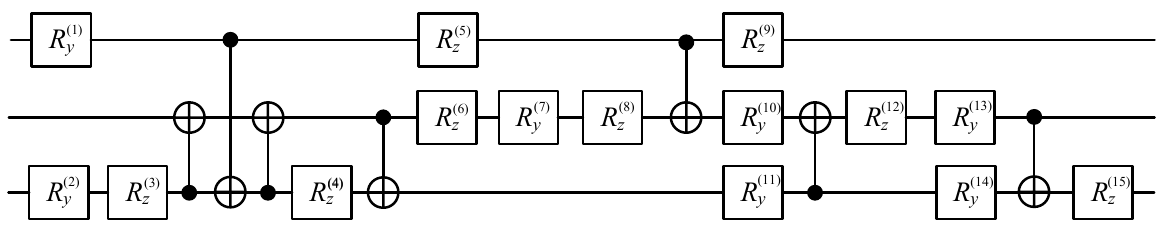}}
\caption{The compiled circuit by VQC based on double $Q$-learning for the three-qubit unitares, \ie three-qubit W-state preparation (WSP3),  Toffoli gate (CCNOT) and  three-qubit quantum Fourier transform (QFT3).}
\label{fig:threeQubitCircuit}
\end{figure}

Fig.~\ref{fig:threeQubitCircuit} shows the structures of the quantum circuits generated by our method for the three-qubit unitaries. The depth of the complied circuits for WSP3, QFT3 and CCNOT are 4, 17 and 19, although the numbers of quantum gates in the circuit are 7, 22 and 25. The rotation angles of the quantum gates in these circuits are illustrated in Table \ref{Tab:par3qubit} in \ref{sec:appendix}.

\begin{figure*}
\centering
\includegraphics[width=0.3\textwidth]{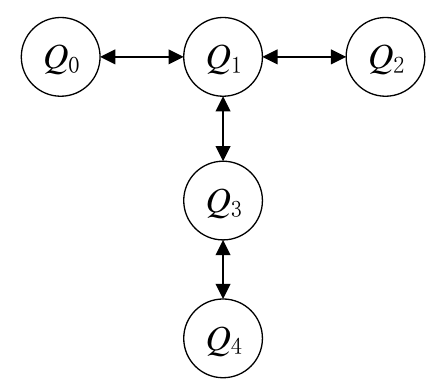}
\caption{\mycomment{The connectivity of IBM's 5-qubit quantum device, \ie `Ibmq$\_$ourense'.}}
\vspace{-10pt}
\label{Fig:ibm}
\end{figure*}

\mycomment{\subsubsection{Simulation considering the connectivity of qubits}
\label{sec:422}
As current available quantum devices do not have completely connected qubits, we also evaluate the performance of the proposed method on the topology of IBM's 5-qubit quantum device, \ie `Ibmq$\_$ourense', as shown in Fig.\,\ref{Fig:ibm}. As WSP3, QFT3 and CCNOT are three-qubit unitaries, we use the first three qubits of `Ibmq$\_$ourense'.
The proposed algorithm can be applied to quantum devices without completely connected qubits by simply removing CNOT$^{i,j}$ gate from the native gate alphabet $\mathcal{A}$ in Algorithm 1 if there is no connection between qubit $i$ and qubit $j$. For `Ibmq$\_$ourense', the native gate alphabet can be denoted by
\begin{eqnarray}
\mathcal{A}_{ibm} = \{&R_z^{0}(\theta), R_z^{1}(\theta), R_z^{2}(\theta), R_y^{0}(\theta), R_y^{1}(\theta), R_y^{2}(\theta),\\
&\text{CNOT}^{01},\text{CNOT}^{10},\text{CNOT}^{12},\text{CNOT}^{21}\},
\end{eqnarray}
where $R_z^{i}$ and $R_y^{i}$ denote the $R_z$ and $R_y$ gates act on qubit $i$. $\text{CNOT}^{ij}$ indicates the CNOT gate acts on qubits $i$ and $j$ where $i$ and $j$ are the control and target qubits, respectively.}

\begin{figure}
\centering
\includegraphics[width=9cm]{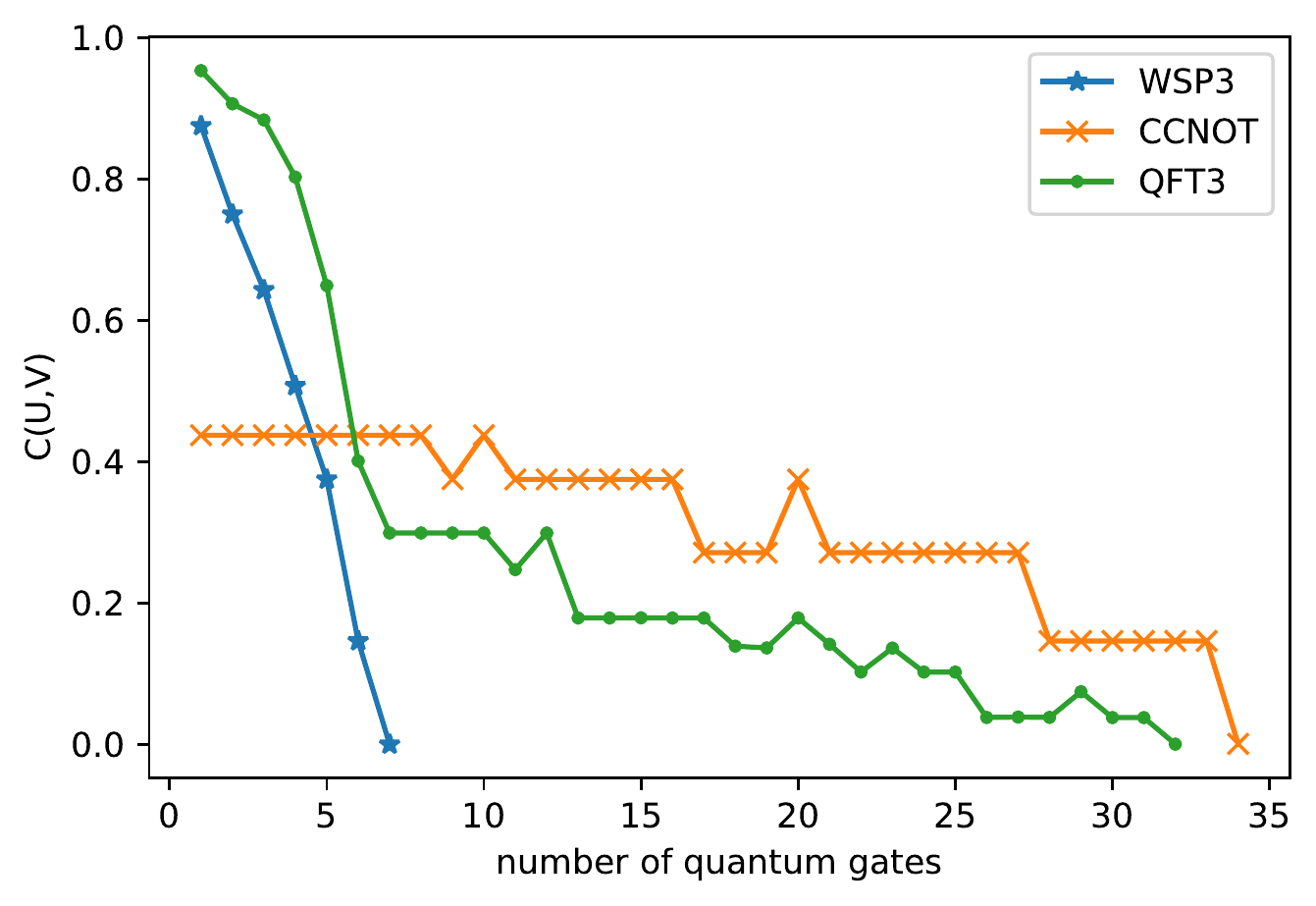}
\caption{\mycomment{The minimum cost achieved by VQC based on double $Q$-learning for the three-qubit unitaries, considering the connectivity of the IBM's 5-qubit quantum device.}}
\label{fig:threeQubitibmq}
\end{figure}

\mycomment{Fig.\,\ref{fig:threeQubitibmq} shows the result of quantum compiling considering the connectivity of IBM's 5-qubit quantum device. Similar to the completely connected case, the cost $C(U,V)$ decreases with the increasing number of quantum gates in the compiled circuit $V$ and converges to 0 for all the target unitaries. However, more quantum gates are needed in CCNOT and QFT3 as $\text{CNOT}^{02}$ and $\text{CNOT}^{20}$ are not available.  For WSP3, the number of quantum gates is the same as the one in the completely connected case as it dose not need $\text{CNOT}^{02}$ and $\text{CNOT}^{20}$ (see Fig.\,\ref{fig:threeQubitCircuit}).}

\mycomment{\subsubsection{Simulation considering the number of CNOT gates}
\label{sec:423}
In this section, we decrease the number of CNOT gates by adding a penalty term to the reward function as described in Eq.\,(\ref{eq:rCNOT}).
The weighted parameter $\lambda$ is set to be 0.1. We consider the connectivity of qubits on IBM's 5-qubit quantum device. The simulation result in Table \ref{Tab:cnotMini} shows that the penalty term in the reward function can reduce the number of CNOT gates in the circuit. However, more gates are needed to get zero cost.}
\begin{table}
	\centering
	\caption{\mycomment{The cost $C(U,V)$, number of CNOT gates and quantum gates of the compiled circuits by the proposed VQC with and without the penalty term on the CNOT gate in the reward function. The simulation considers the connectivity of qubits on IBM's 5-qubit quantum device}}
    \begin{tabular}{ccccc}
    \hline
   &\textcolor{black}{reward function}&\textcolor{black}{$C(U,V)$}&\textcolor{black}{number of CNOT gates} & \textcolor{black}{number of gates} \\
    \hline \multirow{2}{*}{\textcolor{black}{WSP3}} &\textcolor{black}{ without $C_p$}& \textcolor{black}{2.69E-04}& \textcolor{black}{2} &\textcolor{black}{7}\\
    &\textcolor{black}{with $C_p$}&\textcolor{black}{6.27E-05}&\textcolor{black}{2}&\textcolor{black}{8}\\
\hline
    \multirow{2}{*}{\textcolor{black}{CCNOT}} &\textcolor{black}{without $C_p$}&\textcolor{black}{ 9.03E-04}& \textcolor{black}{14}&\textcolor{black}{34}\\
    &\textcolor{black}{with $C_p$}&\textcolor{black}{3.52E-05}&\textcolor{black}{12}&\textcolor{black}{35}\\
\hline
    \multirow{2}{*}{\textcolor{black}{QFT3}} &\textcolor{black}{without $C_p$}& \textcolor{black}{8.40E-04}& \textcolor{black}{9}&\textcolor{black}{32}\\
    &\textcolor{black}{with $C_p$}&\textcolor{black}{6.45E-04}&\textcolor{black}{8}&\textcolor{black}{35}\\
    \hline
    \end{tabular}
\label{Tab:cnotMini}
\end{table}

\subsection{\mycomment{VQC on larger unitaries}}
\begin{figure}
\centering
\includegraphics[width=12cm]{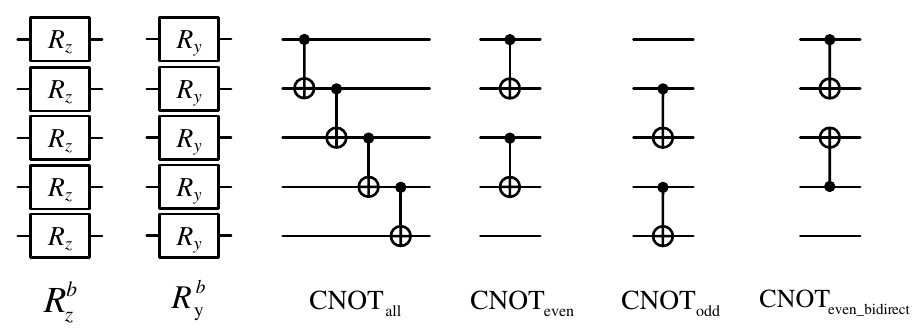}
\caption{\mycomment{Illustrations of common blocks in five-qubit system.}}
\label{fig:LSB}
\end{figure}

\mycomment{In this section, we consider larger unitaries, \ie 4, 5, 6, 7 and 8 qubits. For larger unitaries, Local Hilbert-Schimdt Test is used to measure the difference between the target unitary and the designed unitary. We consider the same target unitary as Ref.\,\cite{KLP19}, \ie
\begin{eqnarray}
U = U_4(\vec \theta')U_3U_2U_1(\vec \theta),
\end{eqnarray}
where $U_1(\vec \theta) = \bigotimes_{i=0}^{n-1}R_z(\theta_i)$,  $U_2 = ...\text{CNOT}_{23}\text{CNOT}_{01}$, $U_3 = ...\text{CNOT}_{34}\text{CNOT}_{12}$ and $U_4(\vec \theta')= \bigotimes_{i=0}^{n-1}R_z(\theta'_i)$. $\text{CNOT}_{ij}$ denotes the CNOT gate where $i$ and $j$ are the control and target qubits, respectively.
For large problem sizes, we can add common blocks to the gate alphabet to decrease the computation complexity.
This strategy can be regarded as a crossover of variable structure ansatze and fixed-template compiling. It can construct the quantum circuits with some common blocks instead of selecting quantum gate one by one.
In this paper, we consider the following blocks:
\begin{eqnarray}
\nonumber R_z^{b} = \bigotimes_{i=0}^{n-1}R_z(\theta_i), R_y^{b} = \bigotimes_{i=0}^{n-1}R_y(\theta_i), \\
\nonumber\text{CNOT}_{all} = ...\text{CNOT}_{12}\text{CNOT}_{01},\\
\nonumber\text{CNOT}_{even} = ...\text{CNOT}_{23}\text{CNOT}_{01},\\
\nonumber\text{CNOT}_{odd} =  ...\text{CNOT}_{34}\text{CNOT}_{12},\\
\text{CNOT}_{even\_bidirect} =  ...\text{CNOT}_{32}\text{CNOT}_{01}.
\end{eqnarray}
We illustrate these blocks on a five-qubit system in Fig.\,\ref{fig:LSB}.}

\mycomment{The learning rate $\alpha$, the discount factor $\gamma$ and the batch size $K$ are set to 0.2, 1 and 128, respectively. The number of quantum circuits generated at each $\epsilon$ is the same as the simulation of three-qubit unitaries. For larger unitaries, local Hilbert-Schimdt Test is used to quantify how close the trained unitary is to the target unitary instead of global Hilbert-Schimdt Test due to the barren plateau phenomena.
Table \ref{Tab:largeScale} shows the local cost achieved by the proposed method for 4, 5, 6, 7 and 8-qubit unitaries. By adding common blocks to the candidate gate alphabet, the local cost can converge to 0 for all the target unitaries with 4 to 8 qubits. We also evaluate the global cost $C(U,V)$ with the quantum circuits generated by $C_{local}(U,V)$
(\ie the same circuit with the same gate parameters). The global costs also converge to 0 for all qubits, which indicates exact compilations of the target unitaries.}
\begin{table}
	\centering
	\caption{\mycomment{The minimum cost achieved by VQC based on double Q-learning for 4, 5, 6, 7 and 8-qubit unitaries.
$C_{local}(U,V)$ denotes the local cost between the compiled circuit and the target unitary.
$C(U,V)$ is the global cost between the target unitary and the compiled circuit generated by $C_{local}(U,V)$ (\ie the same structure and the same gate parameters).} }
    \begin{tabular}{cccc}
    \hline
   \textcolor{black}{number of qubits} &\textcolor{black}{number of gates}&\textcolor{black}{$C_{local}(U,V)$ }& \textcolor{black}{$C(U,V)$}\\
    \hline
    \textcolor{black}{4}&\textcolor{black}{11}&\textcolor{black}{6.58E-04}&\textcolor{black}{9.97E-04}\\
    \textcolor{black}{5}&\textcolor{black}{14}&\textcolor{black}{1.99E-04}&\textcolor{black}{9.01E-04}\\
    \textcolor{black}{6}&\textcolor{black}{17}&\textcolor{black}{5.21E-05}&\textcolor{black}{1.85E-04}\\
    \textcolor{black}{7}&\textcolor{black}{20}&\textcolor{black}{1.02E-04}&\textcolor{black}{2.64E-04}\\
    \textcolor{black}{8}&\textcolor{black}{23}&\textcolor{black}{2.08E-04}&\textcolor{black}{5.68E-04}\\
\hline
    \end{tabular}
\label{Tab:largeScale}
\end{table}

\section{Conclusion}
Circuit design aims to find a proper structure of quantum circuit, which includes selecting which quantum gates,  which qubits they act on and the order of these gates.
It largely affects the performance and is crucial in variational quantum compiling. In this paper, we model the circuit design process as a Markov decision process and design a proper structure of quantum circuit to achieve an exact compilation with double $Q$-learning.
The proposed VQC can automatically generate a high-performance quantum circuit by double $Q$-learning with $\epsilon$-greedy exploration strategy and experience replay without any human expertise and labor.
Another advantage is that the proposed method can minimize the number of quantum gates to achieve an exact compilation, which is especially important for NISQ devices as the algorithm performance degrades with runtime due to decoherence process and gate noise.

Simulation results on two-qubit unitaries (\ie CZ, CH, CS, $XX$(3$\pi/2$) and QFT2) and three-qubit unitaries (\ie WSP3, CCNOT and QFT3) show the satisfying performance of the proposed method. Compared to two typical structures in VQC, \ie alternating-pair ansatz and alternating-pair ansatz, the proposed method can achieve lower cost with much less quantum gates. \mycomment{The proposed method can be easily applied to quantum devices with limited connectivity of qubits. The simulation considering the connectivity of qubits on IBM's quantum device shows that the proposed VQC can make exact compilations of WSP3, CCNOT and QFT3. Considering the low fidelity of two-qubit gates, we motivate the agent to use fewer CNOT gates in the compiled circuit by a penalty term in the reward function. Simulation result shows that it can make an exact compilation with fewer CNOT gates at the cost of more one-qubit gates. Ref. \cite{sharma20} provided rigorous theorems stating that the optimal variational parameters in quantum compiling are resilient to various noises such as measurement noise, gate noise and Pauli channel noise. Hence, the proposed method can be practically useful for noisy intermediate-scale quantum devices.}

\ack This work is supported by Guangdong Basic and Applied Basic Research Foundation (Nos.\,2019A1515011166, 2020A1515011204, 2020B1515020050) and the National Natural Science Foundation of China (Nos.\,61802061, 61972091, 61772565).

\section*{References}

\clearpage
\appendix
\section{The values of gate parameters in the compiled quantum circuit}
\label{sec:appendix}
\begin{table}[hb]
	\centering
	\caption{Rotation angles of the quantum gates with parameters in the compiled quantum circuit for the two-qubit  unitares.}
	\begin{tabular}{c|cc|cc|cc|cc|cc}
		\toprule
		&CS&$\theta$&CH&$\theta$&CZ&$\theta$&$XX(\frac{3\pi}{2})$&$\theta$&QFT2&$\theta$\\
		\hline
		1&$R_z^{(1)}$&3.9267 & $R_z^{(1)}$&-3.1420&$R_z^{(1)}$& 4.7124&$R_z^{(1)}$&-3.1412&$R_z^{(1)}$&-2.3562\\
        2&$R_z^{(2)}$&2.3567 & $R_z^{(2)}$&-3.9270&$R_z^{(2)}$&-4.7122&$R_z^{(2)}$& 3.1415&$R_z^{(2)}$&-4.7124\\
        3&$R_z^{(3)}$&3.9265 & $R_z^{(3)}$&-5.4981&$R_z^{(3)}$& 1.5710&&&$R_z^{(3)}$&3.9269\\
        4&&&&&&&&&$R_z^{(4)}$&-2.3562\\
        5&&&&&&&&&$R_z^{(5)}$&1.5708\\
		\bottomrule
	\end{tabular}
\label{Tab:par2qubit}
\end{table}

\begin{table}[hb]
	\centering
	\caption{Rotation angles of the quantum gates with parameters in the compiled quantum circuit for the three-qubit  unitares.}
	\begin{tabular}{c|cc|cc|cc}
		\toprule
		&WSP3&$\theta$&CCNOT&$\theta$&QFT3&$\theta$\\
		\hline
		1&$R_z^{(1)}$&0.7854 &$R_z^{(1)}$  &5.7317&$R_y^{(1)}$&4.6934\\
        2&$R_z^{(2)}$&-1.2309 &$R_y^{(2)}$ &6.1157&$R_y^{(2)}$&-0.0376\\
        3&$R_z^{(3)}$&3.1416 &$R_z^{(3)}$ &6.0494&$R_z^{(3)}$&2.7728\\
        4&$R_z^{(4)}$&-0.7854 &$R_z^{(4)}$ &3.1418&$R_z^{(4)}$&5.8863\\
        5&$R_y^{(5)}$&3.1416 &$R_z^{(5)}$ &-3.6965&$R_z^{(5)}$&-3.8488\\
        6&& &$R_y^{(6)}$ &3.1416&$R_z^{(6)}$&0.7768\\
        7&& &$R_z^{(7)}$ & -1.8014&$R_y^{(7)}$&-1.6191\\
        8&& &$R_y^{(8)}$ &-4.8526&$R_z^{(8)}$&1.5786\\
        9&& & $R_z^{(9)}$&-3.9271&$R_z^{(9)}$&-1.2562\\
        10&& & $R_y^{(10)}$&-3.1416&$R_y^{(10)}$&-2.3667\\
        11&& &$R_z^{(11)}$ &2.3561&$R_y^{(11)}$&-4.7129\\
        12&& &$R_z^{(12)}$ &3.9271&$R_z^{(12)}$&5.4917\\
        13&& &$R_z^{(13)}$ &-5.4977&$R_y^{(13)}$&6.2535\\
        14&& &$R_z^{(14)}$ &5.4975&$R_y^{(14)}$&5.5126\\
        15&& &$R_y^{(15)}$ &1.6096&$R_z^{(15)}$&-4.7659\\
		\bottomrule
	\end{tabular}
\label{Tab:par3qubit}
\end{table}
\end{document}